%

\def\lsim{\mathrel{\scriptstyle{\buildrel < \over \sim}}}
\magnification 1200
\baselineskip=17pt


\centerline{\bf SUPERCONDUCTIVITY VERSUS TUNNELING} 
\centerline{\bf IN A DOPED ANTIFERROMAGNETIC LADDER}
\vskip 50pt
\centerline{J. P. Rodriguez}
\medskip
\centerline{\it Instituto de Ciencia de Materiales,
Consejo Superior de Investigaciones Cient{\'\i}ficas,}
\centerline{\it 
Cantoblanco, 28049 Madrid, Spain  and}
\centerline{{\it Dept. of Physics and Astronomy,
California State University,
Los Angeles, CA 90032, USA.}\footnote*{Permanent address.}}
\vskip 30pt
\centerline  {\bf  Abstract}
\vskip 8pt\noindent
The low-energy charge excitations of a doped antiferromagnetic ladder
are modeled by a system of interacting spinless fermions that live
on the same ladder.  A relatively large spin gap is assumed
to ``freeze out'' all spin fluctuations.  
We find that the formation of rung hole pairs coincides
with the opening of a single-particle gap for charge excitations 
along  chains
and with the absence of coherent tunneling in between chains.
We also find   that such hole pairs condense into either a crystalline
or superconducting state as a function of the binding energy.
\bigskip
\noindent
PACS Indices: 74.20.Mn, 75.10.Jm, 74.72.Jt, 74.25.Fy
\vfill\eject
One of the strangest features of high-temperature superconductors
is the semiconducting nature of electric transport 
perpendicular to the copper-oxygen
planes common to these materials 
shown in the normal state.$^1$
Anderson and co-workers have proposed that this behavior is
intrinsic, and that it results from the microscopic coincidence
of Cooper pairing with    incoherent tunneling in between adjacent metallic
planes.$^2$  Analogous behavior has been observed recently in the
doped ``ladder'' materials.$^{3}$  
In particular, a large anisotropy 
in the electronic conduction
with respect to the ladder direction  appears$^4$ at compositions that
exhibit superconductivity.$^{5,6}$  The authors of ref. 4 suggest that the 
chain version of the above inter-layer pair/tunneling (IPT)
mechanism is responsible for the phenomenon.$^{7,8}$

Such ``ladder'' materials are made up of a parallel arrangement of
magnetically isolated antiferromagnetic ladders that fill copper-oxygen
planes similar to those of the cuprate superconductors.  They notably
show a spin gap on the order of 
$\Delta_{\rm spin}\sim 500\,{\rm K}$ when undoped,
in accord with theoretical expectations based on the 
spin-1/2 antiferromagnetic Heisenberg ladder.$^3$   The actual ladders in 
these materials can also be doped
with a concentration $x$ of mobile holes,$^{9}$  in which case the spin gap
persists.  Again, theoretical studies of the 
$t-J$ model for a lightly doped  ladder, $tx < J$, 
find evidence for the formation of {\it hole pairs} along rungs
that leave the remaining singlet bonds along  the rungs intact.$^{10}$  
The spin gap persists,  as a result, and
charge excitations are left as the only low-energy excitations.

In this paper, we shall model the  low-energy singlet  charge excitations
of a doped antiferromagnetic ladder  by interacting spinless
fermions that live on the same ladder in the vicinity of
half filling.$^{11}$  We presume, therefore, that spin excitations
are frozen out due to a relatively large spin gap of order
$\Delta_{\rm spin}\sim J/2$. 
A bosonization analysis based on the
corresponding Luther-Emery  model,$^{12-15}$  and generalizations
thereof,$^{16,17}$  yields that the appearance of hole pairs
along rungs coincides with the absence of coherent tunneling in
between chains, as well as with the creation of a gap for
single-particle excitations along  chains.  Also, while the
hole pairs generally crystallize into a charge density-wave (CDW)
state in the weak-coupling limit,$^3$  we predict that they
Bose condense into an IPT-type
superconductor at  large enough binding energies (see Fig. 1).   
Last, the correlations in between pairs that form along the chains
are found to be short range when 
hole pairing along rungs occurs.

To motivate the spinless fermion analysis that will follow,
consider first the standard nearest-neighbor $t-J$ model Hamiltonian 
$$H = \sum_{\langle i,j\rangle} 
[-t(\tilde c_{i,s}^{\dag} \tilde c_{j,s} + {\rm h.c.}) +
 J\,\vec S_{i}\cdot\vec S_{j}]\eqno (1)$$
for a doped antiferromagnetic ladder.                         
Above, $\tilde c_{i,s}^{\dag}$
creates a spin $s$ electron on site $i$ 
as long as this site is unoccupied,
while $\vec S_{i}$  measures the
spin at site $i$.  At half-filling, the $t-J$ model (1)
reduces to the Heisenberg ladder, which is known to
have a spin gap $\Delta_{\rm spin}\cong J/2$.$^{3}$
The latter persists in the presence of a dilute hole
concentration, $tx < J$.$^{10}$
In general, long wavelength electronic 
excitations within chains can be described via the standard Luther-Emery
(LE) model.$^{12-15}$
The spin gap present in lightly doped $t-J$ ladders
indicates then  that the intra-chain LE backscattering term has the form
$$H_{\rm back} \sim  -(\Delta_{\rm spin}/a)\sum_{l=1}^2 \int dx\, {\rm cos}
[\phi_{\uparrow}(x,l) - \phi_{\downarrow}(x,l)]
\eqno (2)$$
where $\phi_s(x,l)$ is the bosonic field that represents
collective particle-hole excitations of spin $s$ electrons
at position $x = i\cdot a$ along chain $l$ of the ladder.$^{18}$
Yet long-wavelength charge excitations
have a typical energy scale on the order of the
kinetic energy, $\Delta_{\rm charge}\sim tx$, which is small
in comparison to the spin gap, $\Delta_{\rm spin}$, in the low
doping limit $tx\ll J$.
We may therefore  take $\Delta_{\rm spin}\rightarrow\infty$ in
this limit.
The bosonic spin degrees of freedom are frozen in such case:
$\phi_{\uparrow} = \phi = \phi_{\downarrow}$.  The remaining
(bosonic) charge degree of freedom $\phi(x,l)$ 
must    then correspond
to an effective spinless fermion on each chain.  
This idea is developed below.

In general, the Hamiltonian for
a system of $N$ consecutively coupled  chains of spinless fermions$^{8}$
can be divided into parallel and perpendicular parts,
$H = H_{\parallel} + H_{\perp}$, where
$$H_{\parallel} = \sum_{l = 1}^N \sum_i [-t_{\parallel}
(f_{i,l}^{\dag} f_{i+1,l} + {\rm h.c.}) +
V_{\parallel} n_{i,l} n_{i+1,l}]\eqno (3)$$
and
$$\eqalignno{H_{\perp} = 
\sum_{l = 1}^N \sum_i[-t_{\perp}
(e^{i\Phi/N} f_{i,l}^{\dag} f_{i,l+1} & +  {\rm h.c.}) +
U_{\perp} n_{i,l} n_{i,l+1} +\cr 
& + V_{\perp} (n_{i,l} n_{i+1,l+1} + n_{i,l+1} n_{i+1,l})]
& (4)\cr}$$
describe  respectively the quantum
mechanics within and in between chains.
Here, $f_{i,l}$ denotes the annihilation operator for the spinless
fermion on the $i^{\rm th}$ site of
chain $l$, with occupation number
$n_{i,l} = f_{i,l}^{\dag} f_{i,l}$.  
Also, $t_{\parallel}$ and $t_{\perp}$ are the
nearest-neighbor hopping matrix elements, while $V_{\parallel}$,
$U_{\perp}$ and $V_{\perp}$ are the model interaction energies.  In the
case of open perpendicular boundary conditions, we shall set 
$f_{i,N+1} = 0 = f_{i,N+1}^{\dag}$.   On the other hand,
the identifications
$f_{i,N+1} = f_{i,1}$ and $f_{i,N+1}^{\dag} = f_{i,1}^{\dag}$
produce periodic boundary conditions, in which case $\Phi$
denotes the magnetic flux along the parallel direction.
Consider now the simple ladder, with $N=2$ chains and open
perpendicular boundary conditions ($\Phi = 0$).
Since low-energy spin excitations are frozen out due to the
formation of singlet bonds along the rungs,$^{3}$  
it is natural to identify the
true electron field $c_{i,l,s}$ with 
the spinless fermion field following
$f_{i,1} = c_{i,1,s(i)}$  and 
$f_{i,2} = c_{i,2,-s(i)}$, where $s(i)$ represents the
antiferromagnetic spin configuration on a given chain.
After some algebraic manipulations, it can be shown that the above
spinless-fermion Hamiltonian takes the form
of an extended Hubbard model$^{14}$ in transverse magnetic
field:
$$\eqalignno{H = \sum_i\Biggl[&-t_{\parallel} \sum_{l=1}^2
(f_{i,l}^{\dag} f_{i+1,l} + {\rm h.c.}) -
t_{\perp}(f_{i,1}^{\dag} f_{i,2} + {\rm h.c.}) +\cr
&+(U_{\perp}+2V_{\perp}) n_{i,1} n_{i,2} - {1\over 2} V (n_{i+1} - n_i)^2 -
{1\over 4} V^{\prime} (m_{i+1} - m_i)^2\Biggr], & (5)\cr}$$
where $n_i = n_{i,1} + n_{i,2}$ and $m_i = n_{i,2} - n_{i,1}$,
and where $V = (V_{\parallel} + V_{\perp})/2$
and $V^{\prime} = V_{\parallel} - V_{\perp}$.
It is important to remark that the interaction terms in this ladder model
are  invariant with respect to $SU(2)$ rotations of the chain labels
if  $V_{\parallel} = V_{\perp}$. 

We now rotate to the bonding-antibonding basis,
$f_{i,\pm} = 2^{-1/2}(f_{i,2} \pm f_{i,1})$,
that diagonalizes the  transverse kinetic energy (5).
In the limit near (but not at) half-filling, all umklapp processes are
negligible.  Taking the continuum limit of the ladder model (5) 
 {\it {\`a} la} Kogut and Susskind$^{14}$ 
then yields the 
Luttinger model $H = H_{\parallel}^{\prime} + H_{\perp}^{\prime}$, where
$$H_{\parallel}^{\prime} =  \sum_{n}\int dx\Biggl[
 2t_{\parallel} a \Bigl(L_n^{\dag}  i\partial_x L_n
 - R_n^{\dag} i\partial_x R_n\Bigr)
 + 4V a L_n^{\dag}R_n^{\dag} R_n L_n 
 - \mu_n (L_n^{\dag} L_n +  R_n^{\dag} R_n) \Biggr],\eqno (6)$$
and 
$$H_{\perp}^{\prime} = H_{\perp ,1}^{\prime} + H_{\perp ,2}^{\prime}
 +  H_{\perp ,4}^{\prime} + H_{\perp ,\rm pair}^{\prime}$$
are rotated parallel and perpendicular pieces,
with a backscattering term
$$\eqalignno{
H_{\perp ,1}^{\prime} = & \sum_{n < n^{\prime}} \int dx 
(U_{\perp}-2V_{\parallel}) a
\Bigl[
L_n^{\dag} R_{n^{\prime}}^{\dag} L_{n^{\prime}} R_n
+ {\rm h.c.}\Bigr], & (7)\cr}$$
with    inter-band    forward scattering terms
$$\eqalignno{
H_{\perp ,2}^{\prime} = & \sum_{n < n^{\prime}} \int dx
2 (V_{\parallel}-V_{\perp}) a
\Bigl[
L_n^{\dag} R_{n^{\prime}}^{\dag} R_{n^{\prime}} L_n
+ {\rm h.c.}\Bigr], & (8)\cr
H_{\perp ,4}^{\prime} = &  \sum_{n < n^{\prime}} \int dx 
(U_{\perp}+2V_{\perp}) a
:(L_n^{\dag} L_n +  R_n^{\dag} R_n):
:(L_{n^{\prime}}^{\dag} L_{n^{\prime}} +  
R_{n^{\prime}}^{\dag} R_{n^{\prime}}):, & (9)\cr}$$
and with an inter-band pseudo-triplet pairing interaction
$$\eqalignno{
H_{\perp ,\rm pair}^{\prime} = & \sum_{n < n^{\prime}} \int dx
2 (V_{\parallel} - V_{\perp}) a
\Bigl[
L_n^{\dag} R_{n}^{\dag} R_{n^{\prime}} L_{n^{\prime}}
+ {\rm h.c.}\Bigr]. & (10)\cr}$$
Here,     
$e^{ik_F x} R_n (x) = L^{-1/2}\sum_k e^{ikx} a_n(k)$ and
$e^{-ik_F x} L_n (x) = L^{-1/2}\sum_k e^{ikx} b_n(k)$
denote field operators for right and left
moving spinless fermions in the bonding or antibonding
band $n = +,-$ 
for  chains of length $L$,
with a Fermi surface at $\pm k_F$.
Above, the symbols `: :' represent
normal ordering.$^{14}$  
Also, $\mu_{\pm} = \pm t_{\perp}$ are the chemical
potentials for each band. 
Notice that Eqs. (6)-(10) describe a Luther-Emery
model for pseudo spin-1/2 fermions.
Since such fermions experience pseudo
spin-charge separation, we have that the coupled chains
factorize following 
$H = H_{\rho} + H_{\sigma}$,
where
$$\eqalignno{
H_{\rho} =  2\pi\hbar v_{\rho} & \sum_{q > 0}\sum_{j = R,L}
\rho_j(q)\rho_j(-q)  + g_{\rho}\sum_q\rho_R(q)\rho_L(-q)
 & (11)\cr
H_{\sigma} =  2\pi \hbar v_{\sigma} & \sum_{q > 0}\sum_{j = R,L} 
\sigma_j(q)\sigma_j(-q) + g_{\sigma}\sum_q\sigma_R(q)\sigma_L(-q) +\cr
& + H_{\perp,1}^{\prime} + H_{\perp,\rm pair}^{\prime}
 - (2L)^{1/2} t_{\perp}[\sigma_L(0)+\sigma_R(0)]
& (12)\cr}$$
are the respective commuting portions of the Hamiltonian.
Here, $\rho_j (q) = 2^{-1/2} [\rho_j(q,+) + \rho_j(q,-)]$
and
$\sigma_j (q) = 2^{-1/2} [\rho_j(q,+) - \rho_j(q,-)]$
are the standard particle-hole operators for total-charge and
pseudo-spin excitations with respect to the bands $n = +, -$,
with $\rho_R (q,n) = L^{-1/2} \sum_k a_n^{\dag}(q+k) a_n(k)$ and
$\rho_L (q,n) = L^{-1/2} \sum_k b_n^{\dag}(q+k) b_n(k)$.
The Fermi velocities and interaction strengths for each
component
are renormalized by the inter-band  forward scattering
processes [Eqs. (8) and (9)]  to
$$\eqalignno{
v_{\rho,\sigma} = &  
a[2 t_{\parallel} \pm (U_{\perp}+2V_{\perp})/2\pi]/\hbar, & (13)\cr
g_{\rho,\sigma} = & a[4V \pm(U_{\perp}+2V_{\parallel})], & (14)\cr}$$
where the $+(-)$ signs above correspond to the $\rho (\sigma)$ label.
We remind the reader that it is assumed throughout that the system
of spinless fermions (5) is near half-filling.

To proceed further, we first note that the pure
Luttinger model (11) for the total-charge excitations
along the ladder corresponds precisely to
the (Efetov-Larkin) hard-core boson  model$^{19}$ for the rung  hole pairs 
that is elaborated ref. 20.
The former is characterized entirely by the Luttinger liquid parameter
$K_{\rho} = (2\pi \hbar v_{\rho} - g_{\rho})^{1/2}
/(2\pi \hbar v_{\rho} + g_{\rho})^{1/2}$
that will re-appear below.
The pseudo-spin piece (12) of the present spinless fermion description
for a doped antiferromagnetic ladder is less trivial, however.
Along the $SU(2)$-invariant line, $V_{\parallel} = V_{\perp}$,
the pairing term $H_{\perp,{\rm pair}}^{\prime}$ in Eq. (12)
is null.
Application of the bosonic representation$^{14,15}$
for the spinless fermions then
reveals that a gap, $\Delta_{\sigma}\neq 0$,
opens in the spectrum of  the
pseudo-spin excitations (12) 
for $g_{\sigma}/a = 2V_{\perp} - U_{\perp} > 0$ in the absence
of transverse hopping,$^{13}$ $t_{\perp} = 0$.
For general inter-chain hopping, $t_{\perp}\neq 0$, and interactions,
$V_{\parallel}\neq V_{\perp}$,
it is instructive to  move
along the Luther-Emery ``line'' $g_{\sigma} = 6\pi \hbar v_{\sigma}/5$,
in which case the spinless fermions that correspond to 
the pseudo-spin system  (12)
are governed by the noninteracting Hamiltonian$^{12}$
$$\eqalignno{
H_{\sigma} =  &  \hbar v_F^{\prime} \sum_k
k(a_k^{\dag} a_k - b_k^{\dag} b_k)
- 2^{1/2}t_{\perp} \sum_k (a_k^{\dag} a_k + b_k^{\dag} b_k)\cr
 &  + \Delta_{\sigma}\sum_k (a_k^{\dag} b_k + {\rm h.c.})
+ \Delta^{\prime}_{\sigma}\sum_k (a_k^{\dag} b_{-k}^{\dag} + {\rm h.c.}).
&  (15)\cr}$$
Here, the pseudo spin gaps have values
$\Delta_{\sigma} = (a/\alpha_0)[(U_{\perp} - 2 V_{\parallel})/2\pi]$
and $\Delta^{\prime}_{\sigma} = (a/\alpha_0)(V^{\prime}/\pi)$,
where $\alpha_0^{-1}$ is the momentum cutoff of the Luttinger model,  
while $v_F^{\prime} = {4\over 5}v_{\sigma}$.
Consider first the $SU(2)$ invariant line $V_{\parallel} = V_{\perp}$,
in which case the gap $\Delta_{\sigma}^{\prime}$ that originates
from the pairing term (10) is null.
The spinless fermions corresponding to the
pseudo-spin sector therefore have energy eigenvalues
$\varepsilon_p = \pm (v_F^{\prime 2}p^2 + \Delta_{\sigma}^2)^{1/2}$.  If 
$N_n$ denotes the number of spinless fermions in band $n$,
then it follows that the band occupations
are equal, $N_+ = N_-$, for $t_{\perp} < \Delta_{\sigma}/2^{1/2}$
and that 
$N_+ - N_- = L\chi_0(t_{\perp}^2 - {1\over 2}\Delta_{\sigma}^2)^{1/2}$
for $t_{\perp} > 2^{-1/2}\Delta_{\sigma}$,$^{21}$ where
$\chi_0 = 2/\pi \hbar v_F^{\prime}$
is the pseudo-spin susceptibility.
Band splitting is therefore {\it absent}
below a  critical inter-chain hopping matrix element.$^{7}$
The general case, $\Delta^{\prime}_{\sigma}\neq 0$,  
off of the $SU(2)$ invariant line
can also be analyzed at the Luther-Emery ``line'' (15).
One finds that the product of all of the energy eigenvalues is 
$\Pi_{p > 0}  (v_{F}^{\prime 2} p^2 + \Delta_{\sigma}^2
+ \Delta^{\prime 2}_{\sigma})^2$, 
which never vanishes!  We conclude
that   the net pseudo spin gap, 
$\Delta_{\rm charge} = (\Delta_{\sigma}^2
+ \Delta^{\prime 2}_{\sigma})^{1/2}$, 
is therefore robust with
respect to $SU(2)$ symmetry breaking in the spinless
fermion model (5).


Yet what is the physical character of the present  ladder model at zero
temperature?   To answer  this question, it is convenient to look again
along the line, $V_{\parallel} = V_{\perp}$, 
in which case the $SU(2)$ non-invariant interaction term in
the model Hamiltonian (5) is absent.  Let us start by assuming
no inter-chain hopping, $t_{\perp} = 0$.  Then the boundary
at $U_{\perp}=2V$ that marks the appearance of 
the pseudo spin gap $\Delta_{\sigma}$
can be identified with the phase boundary that exists between the staggered
CDW state and the rung (hole-pair)  CDW state in the strong-coupling 
limit (see Fig. 1 and ref. 22).  
A self-consistent calculation in terms of the CDW  mean field
$\langle R_n^{\dag} L_n\rangle$ yields the 
approximate formula
$\Delta_{\sigma} = \hbar\omega_0/
{\rm sinh}(\pi\hbar v_{\sigma}/g_{\sigma})$
for the pseudo-spin gap  in the  hole-pair regime,$^{17}$
$g_{\sigma} > 0$,
with prefactor $\omega_0 = v_{\sigma}/\alpha_0$.  
It agrees reasonably well with the previous 
exact result along the Luther-Emery point.
The same     exact analysis 
indicates that band splitting generally  remains absent 
at small enough inter-chain hopping amplitudes
$t_{\perp} <   \Delta_{\sigma}/2^{1/2}$.  
We now address the initial question posed by computing the
correlation functions at long distances and at
long times within this hole-pair regime (see Table I).  
Then $SU(2)$ invariance yields the identity
$\langle f_{0,l}(0) f_{i,l}^{\dag}(t)\rangle =
\langle f_{0,n}(0) f_{i,n}^{\dag}(t)\rangle$
for the intra-chain one-particle propagator, where
$\langle f_{0,n}(0) f_{i,n}^{\dag}(t)\rangle =
G_R(x,t) + G_L(x,t)$ 
is  the propagator in the bonding/anti-bonding
basis ($x = ia$),  with
right and left moving components 
$G_R$ and $G_L$, respectively.
Note that the latter independence of the spinless fermion propagation with
the band index, $n$, is a result of the equal band
occupation, $N_+ = N_-$, present in the hole pair regime,
$t_{\perp} < \Delta_{\sigma}/2^{1/2}$.
The application of the bosonization method  plus
pseudo spin-charge separation yields the forms
$G_{R,L} =  G_{R,L}^{(\rho)} \cdot  G_{R,L}^{(\sigma)}$
for the right and left propagators,
with a Luttinger liquid factor
 $$G_{R,L}^{(\rho)} \sim 
 (x \mp v_{\rho}^{\prime} t)^{-1/2} 
 [2\pi\alpha_0/(x^2 - v_{\rho}^{\prime 2} t^2)^{1/2}]
 ^{\alpha_{\rho}}\eqno (16)$$
due to excitations of the total charge,$^{14}$ 
and with pseudo-spin factor$^{23}$
$$G_{R,L}^{(\sigma)} \sim (x \mp v_{\sigma} t)^{-1/2}
e^{-(\Delta_{\sigma}/\hbar v_{\sigma})(x^2 - v_{\sigma}^2 t^2)^{1/2}}.
\eqno (17)$$
The exponent and velocity that appear in expression (16) have the forms
 $\alpha_{\rho} =  {\rm sinh}^2\psi_{\rho}$
 and   $v_{\rho}^{\prime} = v_{\rho} {\rm sech}\, 2 \psi_{\rho}$,
respectively,  with the hyperbolic   angle $\psi_{\rho}$ set by the relation
 ${\rm tanh}\, 2 \psi_{\rho} = - g_{\rho}/2\pi\hbar v_{\rho}$.
A gap, 
$\Delta_{\rm charge} = \Delta_{\sigma}$, therefore
exists for one-particle charge excitations along chains.
The inter-chain single-particle propagator,
on the other hand, 
is by definition
$\langle f_{0,1}(0) f_{i,2}^{\dag}(t)\rangle =
{1\over 2} [\langle f_{0,+}(0) f_{i,+}^{\dag}(t)\rangle -
\langle f_{0,-}(0) f_{i,-}^{\dag}(t)\rangle]$.  Again,
the band occupations $N_+$ and   $N_-$ are equal
for $t_{\perp} < \Delta_{\sigma}/2^{1/2}$, which implies that 
$\langle f_{0,1}(0) f_{i,2}^{\dag}(t)\rangle$ vanishes at long times
and at long distances in such case.  
In other words, we find that {\it coherent} single-particle
tunneling of charge  
in between chains is entirely suppressed in the hole-pair regime.
Notice that this is consistent with renormalization group  calculations
that find inter-chain hopping to be an irrelevant perturbation
for Luttinger liquids with a pseudo gap in the density of states:$^{24}$
$N(\omega)\propto |\omega|^{\alpha}$, with exponent     $\alpha > 1$.

A similar analysis can be employed to obtain the  static auto-correlators for
various CDW and pair order parameters at long distance.
These results are compiled in Table I.
The intra-chain CDW correlator, for example, has    the form$^{12-15}$ 
$\langle L_l^{\dag}(0) R_l(0) R_l^{\dag}(x) L_l(x)\rangle
\sim {\rm cos}(2 k_F x) (\alpha_0/x)^{K_{\rho}}$. 
It coincides with  the form obtained for the density-density correlator
of the hard-core boson model$^{20}$ for the
rung-hole pairs,$^{19}$ $b_i = f_{i,1} f_{i,2}$.
On the other hand, both  inter-chain CDW order  and
chain-hole-pair autocorrelations are {\it short-range}
in the rung-hole-pair regimes,
with a unique  correlation length 
$\xi_{\sigma} = \hbar v_{\sigma}/2\Delta_{\sigma}$.
Finally, the static autocorrelator for rung-hole      pairs 
 has the asymptotic form
 $\langle L_1(0) R_2(0) R_2^{\dag}(x) L_1^{\dag}(x)\rangle 
 \sim (\alpha_0/x)^{K_{\rho}^{-1}}$.  
This form also coincides with that   obtained from the previously
cited hard-core boson model for the propagation,
$\langle b_i b_j^{\dag}\rangle$, of rung-hole pairs.
In conclusion, the phase-boundary
separating dominant rung-CDW correlations from dominant 
rung-pair autocorrelations is evidently 
determined by the condition $K_{\rho} = 1$  (see Table I).  
Given the absence of coherent tunneling that characterizes
the rung-hole-pair regime in general, we interpret the latter phase
($K_{\rho} > 1$)  as an IPT-type superconductor (see Fig. 1).

To address the question of the transverse conductivity of the ladder
model (5), we shall now compute the transverse
charge stiffness of the triangular
three-leg ladder.$^{8}$  In particular, consider
the corresponding spinless fermion model  [Eqs. (3) and (4), $N=3$]
with a magnetic flux $\Phi$ threading each triangle formed by
the rungs.  Then along the line $V_{\parallel} = V_{\perp}$,
the interaction terms of this model Hamiltonian are invariant
with respect to $SU(3)$ rotations of the chain labels. 
After making  algebraic manipulations similar to those employed to achieve
the form (5) for the simple ladder, we obtain  the form
$$\eqalignno{H = \sum_i\Biggl[&-t_{\parallel} \sum_{l=1}^3
(f_{i,l}^{\dag} f_{i+1,l} + {\rm h.c.}) -
t_{\perp}\sum_{l=1}^3
(e^{i\Phi/3}f_{i,l}^{\dag} f_{i,l+1} + {\rm h.c.}) +\cr
&+(U_{\perp}+2V) \sum_{l=1}^3n_{i,l} n_{i,l+1} 
- {1\over 2} V (n_{i+1} - n_i)^2\Biggr] & (18)\cr}$$
for the triangular ladder model Hamiltonian, 
where $n_i = \sum_{l=1}^3 n_{i,l}$ is
the manifestly $SU(3)$ invariant number operator.
After rotating to the basis 
$f_{i,0} = 3^{-1/2}\sum_{l=1}^3 f_{i,l}$ and
$f_{i,\pm} = 3^{-1/2}\sum_{l=1}^3 e^{\pm i 2\pi l/3} f_{i,l}$
that diagonalizes the transverse kinetic energy, we obtain
the previous Luttinger model  [Eqs. (6), (7) and (9)], 
but with the band index
summed  over the new basis $n = 0,+,-$.  The new chemical
potentials  for each band are $\mu_0 = 2t_{\perp}{\rm cos}(\Phi/3)$
and $\mu_{\pm} = 2t_{\perp}{\rm cos}[(\Phi\pm 2\pi)/3]$.
Now consider the special line $U_{\perp} = -2V$, 
along which the inter-band 
forward scattering interaction (9) vanishes. 
What remains is a generalized backscattering model with three internal 
quantum numbers.$^{16}$
A mean-field analysis of this model$^{17}$ 
finds that long-range CDW order of the type
$\langle R^{\dag}_n L_n\rangle$
is stable for effective attraction between
rungs,  $U_{\perp} < 0$, with a single-particle gap
$\Delta_{\sigma}\cong 2\hbar\omega_0\,e^{-2\pi t_{\parallel}/3|U_{\perp}|}$,
and prefactor $\hbar\omega_0\sim (a/\alpha_0) t_{\parallel}$.   Hence for
small enough inter-chain hopping, $t_{\perp} < \Delta_{\sigma}/2$, the
chemical potential of each band lies within the gap, which
means that the 
transverse charge stiffness, $\partial^2 E_0/\partial \Phi^2 |_{0}$,
is {\it null}.
In addition,
since  the band occupations, $N_0$ and $N_{\pm}$,
are all equal for  $t_{\perp} < \Delta_{\sigma}/2$, the single-particle
intra-band amplitudes $\langle f_{0,n}(0) f_{i,n}^{\dag}(t) \rangle$ 
are then all
equal.  The inter-chain single-particle amplitude
$\langle f_{0,l}(0) f_{i,l+1}^{\dag}(t) \rangle =
{1\over 3}\sum_{n} e^{i2\pi n/3} \langle f_{0,n}(0) f_{i,n}^{\dag}(t) \rangle$
must therefore vanish as well, in agreement with the previous
case of the simple ladder.
On the basis of this mean-field analysis,$^{17}$
we conclude that the present ladder model allows {\it no} coherent transport
whatsoever in between chains in the (rung) hole-pair
regime.

Let us   now apply 
 the spinless fermion model (5) 
to doped antiferromagnetic
ladders  by first drawing a    
comparison  with the corresponding  $t-J$ model.$^{10}$
In the limit $\Delta_{\rm spin}\rightarrow\infty$ that is
assumed throughout due to the relatively small energy scale, $tx$,
for charge excitations, triplet excitations are forbidden.
This means that the only possible charge carriers are hole pairs aligned 
parallel to the either the  rungs or to the chains of the ladder
(see Fig. 2).
The spinless fermion system (5) describes the motion of these
objects.
Since rung hole pairs are responsible for coherent charge transport
along the ladder, 
we have  $t_{\parallel} = t$.
In other words, the hopping of the spinless fermion along chains 
accounts for the effective tunneling of rung hole pairs between rungs.
Yet what is the value of the inter-chain model parameter $t_{\perp}$?
First, observe that the rotation of a rung hole pair into
a chain hole pair is a two-stage process.  As depicted in Fig. 2,
the initial and final singlet pair
 states ($S = 0$) pass through an intermediate 
triplet pair state ($S = 1$) that is the lowest energy spin-excitation
of the system.$^{3,20}$  Second order perturbation theory then yields
the matrix element
$t_{\times} =  t_{\parallel} t_{\perp}/\Delta_{\rm spin}$
 for the rotation of a rung hole pair
into a chain hole pair, and vice-versa.
Yet since such $90^{\circ}$ rotations represent the low-energy
single-particle
charge excitations of the ladder, we have the identity
$t_{\times} = \Delta_{\rm charge}$.  This yields  
the expression 
$t_{\perp} =  (\Delta_{\rm spin}/t_{\parallel})\Delta_{\rm charge}$
for the effective inter-chain hopping matrix element.
Yet since $t_{\parallel} = t$ and $\Delta_{\rm spin} < J/2$, we obtain
the desired inequality
$$t_{\perp}  <  \Delta_{\rm charge}/2 \eqno (19)$$
for $J <  t$.  This indicates that the simple $t-J$ ladder
is indeed consistently within the rung-hole pair
regime  per the spinless fermion description (5).
Note also that (19) implies that coherent motion (11) of rung-hole
pairs  represents the only gapless charge excitation of the
$t-J$ ladder.  This is consistent with exact diagonalization
results$^{25}$
that  find  a   single-particle gap $\Delta_{\rm charge}$ of order 
${3\over 2}tx$    (the energy to {\it add} an electron).$^{20}$
The latter energy scale is small in comparison to the spin gap at low doping, 
$tx \ll J$,
which justifies use of the spinless fermion model (5).

The remaining effective interaction parameters of the
spinless fermion model (5) will be considered to
be phenomenological.
For simplicity, let us move along the $SU(2)$ invariant
line $V_{\parallel} = V_{\perp} > 0$.
We then   notably predict a phase transition in between
a rung-CDW state and an IPT-type  superconductor
at $K_{\rho} = 1$ (see Fig. 1).
Comparison of the corresponding correlation exponents shown in Table I
with  those obtained from a
density-matrix renormalization group analysis of the $t-J$ ladder$^{26}$
indicate that such a doped antiferromagnet 
is in the vicinity of this            superconductor-insulator
transition; i.e., $K_{\rho}\sim 1$.
On this basis, we conclude  that coherent
single particle tunneling in between chains
is absent in a  lightly doped antiferromagnetic ladder,
since $\Delta_{\sigma}\neq 0$ by   Eq. (19).
This does not exclude the possibility of (coherent) Josephson  tunneling of
hole pairs in between adjacent ladders, however.  
To address this issue, consider
two neighboring doped ladders.  
Let us also suppose that adjacent ladders are shifted with respect to 
each other by half a lattice constant, which is in fact the case
for real ladder systems that exhibit superconductivity.$^{3,5,6}$
The dynamics of the rung-hole pairs, $b_i = f_{i,1} f_{i,2}$, 
is then 
equivalent to that of  coupled spin-$1\over 2$ $XXZ$ chains
in magnetic field.$^{19,20}$
The frustrating nature of the ``zig-zag'' (Josephson)
coupling in between chains$^3$ effectively reduces this system to
isolated $XXZ$ chains, each  with a renormalized intra-chain (Josephson)
coupling.$^{27}$  We thus recover the previous superconductor/CDW
transition, but with  $K_{\rho}$ now dependent on the
inter-chain Josephson coupling as well.

Concerning  the experimental situation, the
incoherent tunneling that is characteristic of the 
rung-CDW  phase could explain 
the large conduction anisotropy seen
in the normal state of antiferromagnetic ladder materials.$^{4,9,28}$
In addition, if such a rung-CDW state were to be
pinned, then all components of the resistivity
tensor would exhibit insulating behavior in the
low-temperature limit.  
This is indeed observed experimentally.$^{4}$
It must be pointed out, however, that 
whether or not the low-temperature conductivity
in doped antiferromagnetic ladder materials is
intrinsic remains to be determined (see refs. 4, 9 and 28).
If, on the other hand, the rung-CDW state  would depin at some
elevated temperature, then the generic Drude response 
characteristic of the present ladder model (5) would
yield metallic behavior in the longitudinal resistivity.
The latter is also observed
in antiferromagnetic ladder materials at relatively high 
temperatures.$^{4,28}$
Last, such materials are observed to go superconducting
under extreme presure.$^5$  
A transition under pressure from a rung-CDW state to an IPT superconductor
due to a
 strong dependence in the binding energy,$^{10}$
$-U_{\perp}\sim J$, of the rung-hole pairs  with the lattice constants,
for example, could account for this phenomenon.
Very recently, however, the observation of high-temperature
superconductivity in doped antiferromagnetic ladder materials
at ambient pressure has been reported.$^6$
The above discussion suggests that this system could transit into
a rung-CDW groundstate, on the contrary,
 by appropriate variations in the doping levels, or by varying
other parameters like the pressure.

This work was supported in part by National
Science Foundation grant No. DMR-9322427.
The author thanks G. Gomez-Santos, A. Leggett, D. Poilblanc, 
P. Sacramento, and V. Vieira for discussions.

\vfill\eject
\centerline{\bf References}
\vskip 16 pt

\item {1.} T. Ito et al., Nature (London) {\bf 350}, 596 (1991).

\item {2.} J.M. Wheatley, T.C. Hsu, and P.W. Anderson, Phys. Rev. B
{\bf 37}, 5897 (1988);  S. Chakravarty, A. Sudb\o, P.W. Anderson,
and S.P. Strong, Science {\bf 261}, 337 (1993).


\item {3.} T.M. Rice, Z. Phys. B {\bf 103}, 165 (1997).

\item {4.} N. Motoyama, T. Osafune, T. Kakeshita, H. Eisaki,
and S. Uchida, Phys. Rev. B {\bf 55}, R3386 (1997).

\item {5.} M. Uehara, T. Nagata, J. Akimitsu, H. Takahashi,
N. M\^ ori, and K. Kinoshita, J. Phys. Soc. Jpn. {\bf 65},
2764 (1996).

\item {6.} L. Leonyuk, G.-J. Babonas, R. Szymczak, H. Szymczak,
M. Baran, A. Reza, V. Maltsev, L. Shvanskaya and V. Rybakov,
Europhys. Lett. {\bf 45}, 387 (1999).

\item {7.} D.G. Clarke, S.P. Strong, and P.W. Anderson, Phys. Rev. Lett.
{\bf 72}, 3218 (1994).


\item {8.} S. Capponi, D. Poilblanc, and F. Mila, Phys. Rev. B {\bf 54},
17547 (1996).


\item {9.} T. Osafune, N. Motoyama, H. Eisaki, and S. Uchida,
Phys. Rev. Lett. {\bf 78}, 1980 (1997).

\item {10.} E. Dagotto, J. Riera, and D.J. Scalapino, Phys. Rev. B
{\bf 45}, 5744 (1992); 
H. Tsunetsugu, M. Troyer and T.M. Rice,
Phys. Rev. B {\bf 49}, 16078 (1994). 

\item {11.} J.P. Rodriguez, Bull. Amer. Phys. Soc. {\bf 43}, 790 (1998).

\item {12.} A. Luther and V.J. Emery, Phys. Rev. Lett. {\bf 33}, 589 (1974).


\item {13.} S.T. Chui and P.A. Lee, Phys. Rev. Lett. {\bf 35}, 315 (1975).

\item {14.} V.J. Emery, in {\it Highly Conducting One-dimensional Solids},
ed. by J.T. Devreese, R.P. Evrard and V.E. van Doren
(Plenum Press, New York, 1979).

\item {15.} J. Voit, Rep. Prog. Phys. {\bf 58}, 977 (1995).

\item {16.} J.P. Rodriguez, Europhys. Lett. {\bf 39}, 195 (1997).

\item {17.} J.P. Rodriguez, Phys. Rev. B {\bf 58}, 944 (1998).

\item {18.}  The intra-chain LE backscattering term (2)
appears naturally in the weak-coupling analysis of the  Hubbard ladder
[see   L. Balents and M.P.A. Fisher, Phys. Rev. B {\bf 53}, 12133 (1996)]. 

\item {19.} K.B. Efetov and A.I. Larkin, Zh. Eksp. Teor. Fiz. {\bf 69}, 704
(1975) [Sov. Phys. JETP {\bf 42}, 390 (1976)].

\item {20.} M. Troyer, H. Tsunetsugu, and T.M. Rice,
Phys. Rev. B {\bf 53}, 251 (1996).



\item {21.} 
V.L. Pokrovskii and A.L. Talanov, Zh. Eksp. Teor. Fiz. {\bf 78},
269 (1980) [Sov. Phys. JETP {\bf 51}, 134 (1980)].

\item {22.} J.E. Hirsch, Phys. Rev. Lett. {\bf 53}, 2327 (1984).

\item {23.} J. Voit, J. Phys.: Condens. Matter {\bf 8}, L779 (1996).

\item {24.}  D. Boies, C. Bourbonnais, and A.M.S. Tremblay,
Phys. Rev. Lett. {\bf 74}, 968 (1995).


\item {25.} C.A. Hayward, D. Poilblanc, Phys. Rev. {\bf 53}, 11721 (1996).

\item {26.} C.A. Hayward, D. Poilblanc, R.M. Noack, D.J. Scalapino,
and W. Hanke, Phys. Rev. Lett. {\bf 75}, 926 (1995).

\item {27.} J.P. Rodriguez, P.D. Sacramento, V. Vieira,
unpublished.

\item {28.} T. Osafune, N. Motoyama, H. Eisaki, S. Uchida and S. Tajima,
Phys. Rev. Lett. {\bf 82}, 1313 (1999).




\vfill\eject

\noindent {TABLE I.}  Listed is the correlation exponent $\eta$
obtained via the bosonization technique for various
order parameters, $O(x)$, in the doped spin ladder
model (5); i.e.,
$\langle O(x) O^{\dag}(0)\rangle\propto (\alpha/x)^{\eta}$.
The $SU(2)$ - invariant case is assumed (see ref. 15).
 Below,    the value 
 $\eta = \infty$ indicates short-range order,  while 
 $K_{\rho} = (2\pi \hbar v_{\rho} - g_{\rho})^{1/2}
 /(2\pi\hbar v_{\rho} + g_{\rho})^{1/2}$.

\bigskip\bigskip\bigskip

\vbox{\offinterlineskip
\hrule
\halign{&\vrule#&
  \strut\quad\hfil#\hfil\quad\cr
height2pt&\omit&&\omit&&\omit&&\omit&\cr
&Order\hfil&&Order Parameter\hfil&&$\eta$\quad(Staggered CDW)\hfil
&&$\eta$\quad(Rung Hole Pairs) &\cr
height2pt&\omit&&\omit&&\omit&&\omit&\cr 
\noalign{\hrule}
height2pt&\omit&&\omit&&\omit&&\omit&\cr 
&CDW, $\parallel$&&$R_l^{\dag} L_l$ 
&&$K_{\rho} + 1$&&$K_{\rho}$&\cr
&Pair, $\parallel$&&$R_l L_l$&&$K_{\rho}^{-1} + 1$&&$\infty$&\cr
&CDW, $\perp$&&$R_1^{\dag} L_2$
&&$K_{\rho} + 1$&&$\infty$&\cr
&Pair, $\perp$&&$R_1 L_2$&&$K_{\rho}^{-1} + 1$&&$K_{\rho}^{-1}$&\cr
&CDW$^2$&&$R_1^{\dag} R_2^{\dag} L_2 L_1$ 
&&$4 K_{\rho}$&&$4 K_{\rho}$&\cr 
&Pair$^2$&&$R_1 R_2 L_2 L_1$ 
&&$4K_{\rho}^{-1}$&&$4K_{\rho}^{-1}$&\cr  
height2pt&\omit&&\omit&&\omit&&\omit&\cr}
\hrule}


\vfill\eject

\centerline{\bf Figure Captions}
\vskip 20pt
\noindent {FIGURE 1.}   The phase diagram of  the $SU(2)$ - invariant
model (5) for the charge excitations of a doped antiferromagnetic
ladder is displayed in the regime of effective repulsion
within each chain, $V_{\parallel} = V_{\perp} > 0$,
and in  the absence of hopping in between chains.
 The rung hole-pair regime ($\Delta_{\sigma} > 0$) 
 is, nevertheless,  expected to persist for 
 small enough inter-chain hopping  matrix elements
 $t_{\perp} \lsim \Delta_{\sigma}/2$.
In such case, the line separating dominant CDW correlations
from superconducting ones among the rung-hole pairs
is determined by the condition $K_{\rho} = 1$.

\bigskip\bigskip

\noindent {FIGURE 2.} A diagrammatic 
representation is given  for the second-order
matrix element ($t_{\times}$) that connects rung and chain hole pairs.
The spin gap ($\Delta_{\rm spin}$) is presumed to be large in
comparison to any difference in the binding energy ($\varepsilon_{\rm pair}$)
between rung and chain hole pairs.

\end